# Data Quality in Empirical Software Engineering: A Targeted Review

Michael Franklin Bosu and Stephen G. MacDonell

*SERL, School of Computing and Mathematical Sciences*
*AUT University, Auckland 1142 New Zealand*
*michael.bosu@aut.ac.nz, stephen.macdonell@aut.ac.nz*

## ABSTRACT

***Context:*** *The utility of prediction models in empirical software engineering (ESE) is heavily reliant on the quality of the data used in building those models. Several data quality challenges such as noise, incompleteness, outliers and duplicate data points may be relevant in this regard.* ***Objective:*** *We investigate the reporting of three potentially influential elements of data quality in ESE studies: data collection, data pre-processing, and the identification of data quality issues. This enables us to establish how researchers view the topic of data quality and the mechanisms that are being used to address it. Greater awareness of data quality should inform both the sound conduct of ESE research and the robust practice of ESE data collection and processing.* ***Method:*** *We performed a targeted literature review of empirical software engineering studies covering the period January 2007 to September 2012. A total of 221 relevant studies met our inclusion criteria and were characterized in terms of their consideration and treatment of data quality.* ***Results:*** *We obtained useful insights as to how the ESE community considers these three elements of data quality. Only 23 of these 221 studies reported on all three elements of data quality considered in this paper.* ***Conclusion:*** *The reporting of data collection procedures is not documented consistently in ESE studies. It will be useful if data collection challenges are reported in order to improve our understanding of why there are problems with software engineering data sets and the models developed from them. More generally, data quality should be given far greater attention by the community. The improvement of data sets through enhanced data collection, pre-processing and quality assessment should lead to more reliable prediction models, thus improving the practice of software engineering.*

**Keywords:** data quality; data sets; empirical software engineering; literature review.

## 1. INTRODUCTION AND BACKGROUND

Empirical software engineering (ESE) leverages observed data to model and understand software engineering phenomena. Models constructed for effort/cost estimation and defect prediction, based predominantly on quantitative data, have been widely investigated in ESE research. One of the factors that should encourage increased practitioner adoption of such models is the high(er) quality of the data used in their development.

The quality of ESE data sets has therefore come under increasingly close scrutiny in the last decade. Gray et al. [3] identified several data quality challenges with the widely accessible NASA Metrics Program data sets that are often used in defect prediction. The issues include redundant data, inconsistencies, constant attribute values, missing values, and noise. Khoshgoftaar and colleagues have been very active researchers on various data quality issues in ESE, especially regarding noise and missing or incomplete data. In two studies [5, 7] they addressed the issue of missingness through the introduction of imputation techniques. Several noise handling mechanisms were also applied by Khoshgoftaar et al. [4, 8, 9]. Another data quality challenge that has been reported often in the literature is the presence of outliers in ESE data sets [1, 12, 16]. Liebchen & Shepperd [13] reported a systematic review in 2008 and concluded that the issue of data quality had not been given desired attention by the software engineering community, as they found just 23 studies that had highlighted data quality.

The goal of evidence based software engineering as espoused by Kitchenham et al. [11] is to provide the means by which the current best evidence from research can be integrated with practical experience and human values in the decision making process regarding the development and maintenance of software. With this overall intent in mind we performed a targeted literature review covering the period January 2007 to September 2012, to identify evidence of data collection reporting, data pre-processing and data quality issue identification in ESE studies. We believe this review will be useful to both practitioners and researchers. Practitioners will be informed of the state of research and be conversant with (and help prevent) the problems that plague ESE data sets. It offers researchers the opportunity to know the extent of research with respect to data quality and to identify gaps for further research.

We report on the results of our research with a particular emphasis on whether there are associations between these three factors of data quality; data collection, data pre-processing and the identification of data quality problems.

This paper is organized as follows: Section 2 describes the review process, Section 3 reports the review results, Section 4 comprises our discussion of the findings of the study, and Section 5 concludes our study.

## 2. THE REVIEW PROCESS

While our review was deliberately targeted rather than systematic in the true sense, we still developed a protocol for the search, inclusion extraction and evaluation of studies. We refer to this review as targeted because of the limited number of years covered and the journals and conference proceedings selected.

### 2.1 Inclusion Criteria and Research Questions

Studies included in our review must have been designed to estimate, predict or model some aspect of software engineering phenomena, such as effort/cost estimation and defect prediction. In addition, studies introducing measurement programmes or systems were also included, as were studies analyzing or evaluating some aspect of ESE data sets or data quality. Studies that provided comment on previous research were excluded along with comparative studies that did not validate their results using ESE data sets. Similarly, studies that used only artificial data sets as well as studies based on expert judgement were excluded.

This review sought to answer the following primary research questions RQ1-RQ7:

- RQ1: Do ESE researchers assess the quality of data sets prior to their use in modeling?
- RQ2: What is the dominant data quality issue?
- RQ3: How often are data sets pre-processed prior to modeling?
- RQ4: Are data collection procedures reported in ESE papers?
- RQ5: Is data collection reporting associated with data pre-processing?
- RQ6: Is data collection reporting associated with data quality issue identification?
- RQ7: Is data pre-processing associated with data quality issue identification?

We further derived a secondary question RQ8 based on the extracted central theme of the papers with a view to ascertaining whether this might be associated with consideration of our three major elements of data quality (data collection reporting, data pre-processing and data quality issue identification):

- RQ8: Is a study's research theme associated with the three elements of quality under consideration?

### 2.2 Search Strategy

We searched for papers based on an issue-by-issue, manual reading of titles and abstracts of all published papers from selected journals and conferences, an approach proposed by Jørgensen and Shepperd [6] as being appropriate when seeking coverage completeness. We then read thoroughly these potential candidates to decide whether or not to include them in our review. We settled on a total of 221 studies. The choice of publication venues considered in the review was based on previous studies [2, 6, 10, 13, 14] that found most relevant papers from these journals and conferences. The review period was chosen as January 2007 to September 2012. (Thus, note that the 2012 results are incomplete.) Table 1 shows the total numbers of papers selected from each journal or conference.[1]

**Table 1. Journal/Conference Papers in Review**

| Journal/Conference | Number |
|---|---|
| Empirical Software Engineering | 31 |
| Journal of Systems & Software | 20 |
| Information & Software Technology | 19 |
| IEEE Transactions on Software Engineering | 11 |
| Software Quality Journal | 11 |
| IEEE Software | 3 |
| *Total from Journals* | *95* |
| International Symposium on Empirical Software Engineering and Measurement (ESEM) | 63 |
| International Workshop/Conference on Predictor Models in Software Engineering (PROMISE) | 53 |
| International Conference on Evaluation and Assessment in Software Engineering (EASE) | 10 |
| *Total from Conferences* | *126* |
| ***Grand Total*** | ***221*** |

### 2.3 Classification of Papers

In order to answer our research questions we classified our studies according to several categories: whether data collection procedures were reported, whether data were pre-processed, and whether data quality was assessed prior to use in modeling. The central theme of each study was also noted (being one of effort estimation, software quality (commonly, but not exclusively, defect prediction), measurement programme/system, and data quality). In addition we categorized the studies into two groups according to the accessibility of the data sets used (being public or private). Any data set that any researcher can have access to, be it in a repository or reported in a published paper or that can be extracted from a version control system or some other source, is termed public. Private data sets refer to all such data sets that a researcher cannot easily access to replicate a study.

---

[1] The list of papers considered in this review is accessible from: http://tinyurl.com/EASE13-DataQuality

## 2.4 Review Process Threats to Validity

The major threat to the validity of our review arises from the chance that relevant and important studies have been missed, bringing our results into question. This may have occurred due to the limited number of years covered by the study, the limited number of publication venues considered, and the conscious or accidental exclusion of relevant studies. Although this study considered papers published only between 2007 and 2012, we selected venues based on previous studies [2, 6, 10, 13, 14] in order to source from the most important outlets for empirical software engineering papers. We also adopted the approach of Jørgensen & Shepperd [6] by performing an issue-by-issue manual search to increase the potential of including all studies that satisfied our inclusion criteria. We identified 221 papers from both journals and conferences and, while this number is limited by the constraints we imposed in adopting a targeted approach, we believe this volume of studies has the breadth and depth to provide a reasonable and up-to-date assessment of the research community's consideration and treatment of ESE data quality. The techniques, data sets and modes of reporting empirical software engineering research have not changed dramatically from the past (before 2007), and it is our contention that the findings reflect current industry and research practice in dealing with issues of ESE data quality. Executing the data extraction process was also time-intensive because the information we sought was not systematically presented in most of the papers.

Some of the papers under consideration used multiple data sets in which, for instance, data collection reporting was provided for some of the data sets but not others. In such a situation we classified the study as having reported data collection, because we deemed the researcher(s) to be aware of the need for reporting such information. There were very few studies that had this issue. This same approach to classification was applied to the other elements of pre-processing and data quality issue identification.

## 3. RESULTS

We evaluated 221 papers comprising 95 journal papers and 126 papers drawn from conference proceedings. Figure 1 illustrates the number of selected papers published per year in our review. (As noted, the 2012 result is preliminary as data for that year is incomplete.) The majority of studies in the review used public data sets (66%) while the remainder used private data sets.

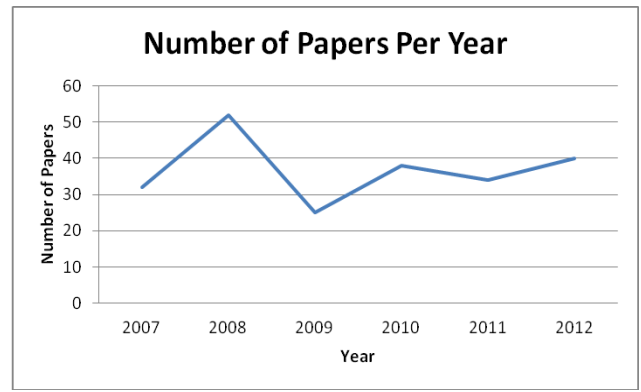

**Figure 1. Number of papers per year**

Figure 2 shows the number of studies that reported on each of the three data quality elements under consideration here. The reporting of data collection procedures was particularly prominent in 2012, in spite of the records for this year being incomplete. In contrast, 2009 was a year in which low numbers of studies addressed data quality in any way.

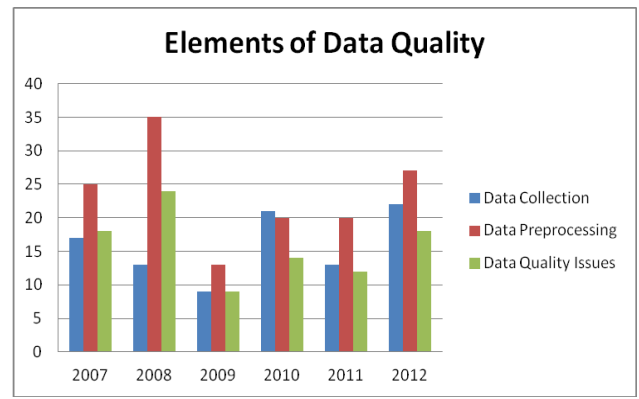

**Figure 2. Data quality reporting trends**

A visual interpretation of Figure 2 implies, among other things, that the more that pre-processing was addressed, the more data quality issues were also identified. There seems to be no identifiable trend, however, between the reporting of data collection procedures and the other two elements being considered, according to Figure 2. Each of these potential associations is investigated in the following subsections.

## 3.1 Data Quality Assessment

- RQ1: Do ESE researchers assess the quality of data sets prior to their use in modeling?

More than half of the reviewed studies (126 of 221, or 57%) did not report on the quality of the data sets being used; rather, the data was used as obtained for the chosen modeling task. The remaining studies (95, or 43%) raised possible issues of data set quality, highlighting the quality challenges encountered and in some cases describing the steps taken to resolve them.

Fifty-four of the papers that highlighted the issue of data quality used data sets from the PROMISE and International Software Benchmarking Standards Group (ISBSG) repositories. Eleven studies relied solely on ISBSG data sets and 23 studies used ISBSG data in combination with data from the PROMISE repository. The remaining 20 studies employed data sets from the PROMISE repository only. The PROMISE and ISBSG repositories have become two of the most widely used sources of data sets in empirical software engineering studies in the last decade. That said, the more recent consideration of data sets drawn from open source software repositories is on the rise. There are data quality challenges associated with all of these sources, as detailed in some of the studies covered for this review. While it is incumbent upon researchers to use secondary data judiciously, it is clear from the above that in many studies (more than half, in our case) data quality is not given any explicit attention. Thus it would be ideal if repositories are populated as far as possible with clean, validated data to reduce the effort expended by researchers in addressing quality issues (if indeed they can) and also to improve the reliability of models generated from the data.

### 3.2 Dominant Data Quality Issue

- RQ2: What is the dominant data quality issue?

The most frequently identified data quality issues are missingness (incompleteness) and outliers, these being noted in 42% and 33% of studies, respectively. This might be due to the fact that there are techniques that easily detect missingness and outliers as compared to, for instance, noise. Another reason might be that these are the predominant problems with software engineering data sets.

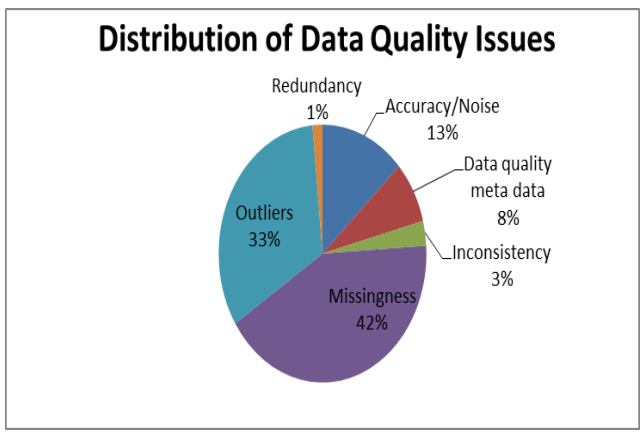

**Figure 3. Distribution of Data Quality Issues**

Figure 3 depicts the distribution of data quality issues identified in the reviewed studies. Noise/accuracy (13%), data quality metadata (8%), inconsistency (3%) and redundancy (1%) comprise the remainder of the distribution. Data quality metadata is noted only in relation to use of the ISBSG repository, so for this to constitute 8% of the reported data quality issues gives credence to the importance of the ISBSG repository in the empirical software engineering research community.

### 3.3 Frequency of Data set Pre-processing

- RQ3: How often are data sets pre-processed prior to modeling?

Pre-processing is reported to be a relatively common practice, with 63% of the reviewed studies noting pre-processing of the data prior to its use. The form of pre-processing varied widely and included: 1) the removal of outliers and/or missing data; 2) the selection (subset selection) and possible reduction of features through the application of techniques such as principal component analysis, curvilinear component analysis, filtering, log filtering, stepwise subset selection, and row and column pruning; 3) normalization of data sets; 4) oversampling and undersampling to balance data sets; 5) transformation of data distributions through the application of log transform, square root transform and lognormal distributions to address skew and kurtosis; and 6) discretization of continuous data into discrete data bins to enable the use of particular analysis methods.

It can be deduced from the above that software engineering data sets are frequently not in a format appropriate for many modeling tasks, but can be pre-processed – with the possible consequence of data loss or error – to be appropriate for specific analysis and modeling methods.

### 3.4 Reporting of Data Collection Procedures

- RQ4: Are data collection procedures reported in ESE papers?

Ninety-five studies (representing 43% of those reviewed) reported on their data collection procedures, although the extent of the description varied substantially, particularly in respect to whether the collection procedure could be replicated. The remaining studies (57%) did not describe how the data sets used in those studies were collected. Most studies that did not report on their data collection procedures used data from the PROMISE and ISBSG repositories; in these cases a description of the attributes or variables being used in the modeling was normally provided.

A total of 69 studies extracted data from version control systems (CVS), defect tracking systems, configuration management systems, source code repositories, issue tracking systems and the like. Of these studies, 65 reported their data collection procedures, representing 73% of all studies that reported their data collection in the review. Data collection reporting therefore appears to be a majority practice for studies that extract data from version control systems or similar.

Data collection reporting for studies utilizing a CVS typically includes information about the tool that was used to extract the data, the name of the system or repository from which the data was drawn, and the metrics or variables extracted.

## 3.5 Data Collection Reporting and Data Pre-processing

- RQ5: Is data collection reporting associated with data pre-processing?

We wanted to find out if there is an association between the reporting of data collection procedures and the pre-processing of data. This basically reflects an assumption that interest in or awareness of the processes needed to collect the data might lead to recognition of the need to pre-process the data prior to modeling. We conducted a Chi-square test of independence to establish this based on the data depicted in Table 2. The null hypothesis and the alternate hypothesis are given below:

$H_o$: Data collection reporting is independent of data pre-processing.

$H_a$: Data collection reporting is associated with data pre-processing.

**Table 2. Data Collection Reporting and Pre-processing**

|  | Data Pre-processed | No Data Pre-processing |
|---|---|---|
| Data Collection reported | 56 | 39 |
| No report on Data Collection | 84 | 42 |

We used a statistical significance level of 0.05 and degrees of freedom of 1. We obtained a Chi-square value of 1.39 and a p-value of 0.238. In light of our result, with the p-value greater than 0.05 (0.238 > 0.05), we fail to reject the null hypothesis. We therefore conclude that data collection reporting is independent the pre-processing of data.

## 3.6 Data Collection Reporting and Data Quality Issue Identification

- RQ6: Is data collection reporting associated with data quality issue identification?

In order to determine whether studies that reported on data collection also identified data quality issues in their data sets we conducted a Chi-square test of independence, using the data shown in Table 3. The null hypothesis and the alternate hypothesis are given below:

$H_o$: Data collection reporting is independent of data quality issue identification.

$H_a$: Data collection reporting is associated with data quality issue identification.

We again used a significance level of 0.05 and degrees of freedom of 1. We obtained a Chi-square value of 5.339 and a p-value of 0.02. Since the p-value is less than 0.05 (0.02 < 0.05), we reject the null hypothesis and accept the alternate, concluding that there is a statistically significant association between data collection reporting and data quality issue identification.

**Table 3. Data Collection Reporting and Data Quality Issues**

|  | Data Quality Issues | No Data Quality Issues |
|---|---|---|
| Data Collection reported | 32 | 63 |
| No report on Data Collection | 62 | 64 |

## 3.7 Data Pre-processing and Data Quality Issue Identification

- RQ7: Is data pre-processing associated with data quality issue identification?

In order to establish whether there is a relationship between studies that pre-processed data and the identification of data quality issues in these papers, we conducted a Chi-square test of independence based on the data in Table 4.

$H_o$: Data pre-processing is independent of data quality issue identification.

$H_a$: Data pre-processing is associated with data quality issue identification.

**Table 4. Data Pre-processing and Data Quality Issues**

|  | Data Quality Issues | No Data Quality Issues |
|---|---|---|
| Data Pre-processed | 75 | 65 |
| No Data Pre-processing | 19 | 62 |

We used a significance level of 0.05 and degrees of freedom of 1. We obtained a Chi-square value of 19.038 and a p-value less than 0.00. Since the p-value is less than 0.05 (0.00 < 0.05), we reject the null hypothesis and accept the alternate hypothesis, and so conclude that there is a statistically significant association between data pre-processing and data quality issue identification. This confirms the result determined visually from Figure 2.

## 3.8 Theme of Study and Elements of Quality

- RQ8: Is a study's research theme associated with the three elements of quality under consideration?

We derived four classes for the central theme of the studies that we reviewed. The Data Quality class encompassed papers that have their major theme as assessing or improving the quality of ESE data sets. For instance, studies that introduced imputation techniques or noise handling techniques were included under this theme. We grouped all papers that estimated effort, schedule, duration and size under Effort Estimation. Studies that introduced measurement programs or systems were grouped under Measurement Programme/System. Papers that addressed the improvement of software quality, via defect prediction, fault estimation and the like, were all grouped under Software Quality.

The number of papers in each theme class is shown in Table 5.

**Table 5. Classification by Theme of Papers**

| Theme of Paper | Number |
|---|---|
| Data Quality | 12 |
| Effort Estimation | 86 |
| Measurement Programme/System | 4 |
| Software Quality | 119 |

None of the twelve papers whose central theme was Data Quality reported how the data to be analyzed was collected. Ten of these papers applied pre-processing methods before assessing the quality of the data: four papers addressed missingness and another four, outliers; one addressed noise; a further paper dealt with several quality issues [3]; one considered the treatment of duplicate data; and one employed the use of metadata quality criteria from the ISBSG repository.

Only eighteen of the 86 papers (or 21%) on Effort Estimation or similar reported their data collection procedures, meaning 68 studies provided no information regarding how the data was collected. Of those papers that did not report, 59 used data sets drawn primarily from the PROMISE and ISBSG repositories. The data sets in these two repositories are secondary data, accounting to at least some degree for the fact that researchers do not report how the data was acquired. That said, the wide availability of such data from a third party does not absolve the researcher of their responsibility to consider and take account of the reliability of its acquisition. While some documentation is provided concerning data collection for some elements of these repositories it is neither comprehensive nor consistent. Sixty-five of the 86 Effort Estimation papers (76%) carried out some form of pre-processing of the data prior to its use for modeling, and again the majority of these studies utilized data drawn from the PROMISE and ISBSG repositories – this provides ample evidence that researchers are aware of the unsuitability of the 'raw' data – that is, as provided or extracted – for modeling, and as such changes are required to adapt them for the task at hand. Fifty-seven of the Effort Estimation papers (66%) raised at least one issue of concern regarding the quality of the data sets, implying that around two-thirds of researchers working on this theme are aware of some of the data quality challenges associated with the use of data drawn from publicly accessible repositories.

All four papers that addressed Measurement Programmes/Systems reported on associated data collection procedures. Three applied some level of pre-processing and two of the four studies raised issues of data quality.

A total of 119 papers were classified under the theme of Software Quality. Seventy-two of these studies (61%) reported on their data collection procedures. Sixty-four studies (54%) performed pre-processing of the data, but only 24 studies (20%) raised concerns about issues of data quality. All but two of the 72 papers that reported on their data collection procedures extracted data from version control systems, defect tracking systems, issue tracking systems, configuration management systems, source code repositories or similar. Sixty-seven of these 72 studies did not raise any data quality concerns regarding their data sets – this could be attributed to an assumption that since the researchers extracted the data themselves, they assumed it to be correct. This is, however, contrary to the quality problem expressed by Mizuno et al. [15] regarding the difficulty of keeping track of all faults in version control systems. In general it appears that data quality consideration is a minority practice among studies that utilize data from source code repositories, version control systems and so on.

## 4. DISCUSSION

Just 23 studies, equating to around 10% of those reviewed, reported on all three elements of data quality that comprise the themes of this study. Fewer than half of the studies (43%) reported any consideration of data quality, confirming the Liebchen & Shepperd [13] finding that data quality consideration is a minority practice in the ESE research community. The majority of the studies that highlighted issues of data quality used data sets drawn from the PROMISE and ISBSG repositories, signifying that there is a degree of acknowledgement among researchers of the data quality challenges that can arise with the use of publicly accessible repositories. While data incompleteness was the most frequently identified quality issue in ESE data sets, most studies did not address the cause(s) of incompleteness (or at least this was not discussed). Awareness of the cause of incompleteness is the first step to reducing its occurrence. Knowing the cause has the potential to lead to the development of suitable preventative measures as well as the use of appropriate imputation techniques to ameliorate this situation.

Several pre-processing techniques are regularly applied to the data sets employed in ESE studies. Close to two-thirds of the papers reviewed (63%) reported some form of pre-processing being applied to the data. The reasons for pre-processing were either to improve the general quality of the data sets or to transform the data into an appropriate format to suit the parameters of the modeling task at hand.

Data collection reporting was a minority practice among the studies reviewed here. For those that did report on data collection procedures, most tended to describe how the data was collected, but not the problems associated with data collection, making it difficult to attribute data quality issues to data collection. Data collection reporting is a predominant practice in studies that extracted data from version control systems, issue tracking systems and so on. In contrast, there was no data collection reporting in the studies that used the PROMISE and ISBSG repositories – this can perhaps be attributed to the fact that these ESE researchers are working with secondary data and as such are not directly involved in the data collection process.

We found no association between the reporting of data collection procedures and studies that pre-processed the data set(s). In contrast, data collection reporting is associated with the reporting of data quality issues in the papers under review. Our review also identified an association between data pre-processing and papers that identified data quality issues.

## 5. CONCLUSION

This review has revealed that data pre-processing has received considerable attention in the ESE research community. The same cannot be said regarding data collection procedures and the identification of data quality issues. Both aspects need to be given greater attention by the ESE community. Problems encountered during data collection and use need to be reported to stimulate research into finding ways of addressing these problems. The fact that a third of the studies reviewed employed private data sets is also of concern for the practice of software engineering, since this signals that there is still a substantial number of studies that cannot be replicated to help improve the practice of software engineering. This is problematic in that replication is a key requirement of sound science. Increasing the public availability of data sets that have also been enhanced through improved collection, cleaning and transformation procedures will lead to more reliable predictive models and thus improve software engineering practice.